\documentclass[aps,prl,showpacs,superscriptaddress,groupedaddress]{revtex4}  
\usepackage{graphicx}  
\usepackage{dcolumn}   
\usepackage{bm}        
\usepackage{amssymb}   
\usepackage{graphicx}
\usepackage{color}
\usepackage{lscape}
\usepackage{amsmath}
\usepackage{epstopdf}
\begin{document}
\title{\LARGE{Comment on "Analytic model of the energy spectrum of a graphene quantum dot in a perpendicular magnetic field"}}
\date{\today}
\author{Babatunde J. Falaye}
\email[E-mail: ]{fbjames11@physicist.net}
\affiliation{Departamento de F\'isica,  Escuela Superior de F\'isica y Matem\'aticas,  Instituto Polit\'ecnico Nacional,  Edificio 9,  Unidad Profesional ALM,  Mexico D.F. 07738,  Mexico}
\author{Guo-Hua Sun}
\email[E-mail:]{sunghdb@yahoo.com}
\affiliation{Catedr\'{a}tica CONACyT,  CIC,  Instituto Polit\'ecnico Nacional,  Unidad Profesional ALM,  Mexico D. F. 07738,  Mexico}
\author{Wen-Chao Qiang}
\email[E-mail:]{qwcqj@163.com}
\affiliation{Faculty of Science,  Xi'an University of Architecture and Technology,
Xi'an 710055,  P.R. China}
\author{Shi-Hai Dong}
\email[E-mail: ]{dongsh2@yahoo.com}
\affiliation{CIDETEC,  Instituto Polit\'{e}cnico Nacional,  Unidad Profesional ALM,  M\'{e}xico D. F. 07700,  Mexico.}


\date{\today}

\maketitle

In recent work by Schnez et al. [PRB 78,  195427 (2008)],  they studied the analytical model of the energy spectrum of a graphene quantum dot in a perpendicular magnetic field. In this comment we first point out that the results Eqs.(5), (6) and (11) presented by them in [1] are not reliable and then give our results.

The energy spectrum of graphene is linear at two inequivalent points in the Brillouin zone. The Hamiltonian reads
\begin{equation}
H=v_{F}(\vec{p}+e\vec{A})\cdot\vec{\sigma}+\tau V(x, y)\sigma_z,
\end{equation}
where $v_{F}$ is Fermi velocity and $\tau=\pm1$ distinguishes the two valleys $K'$ and $K$ and $\vec{\sigma}=(\sigma_x, \sigma_y)$ are Pauli's spin matrices.

Using the infinite-mass boundary,  i.e.,  one has $V(r)=0$ for $r\leq R$ and $V(r)=\infty$ for $r>R$. Thus for $V(r)=0$,  we can find the eigenvalue of the problem via $H\Psi=E\Psi$,  where $\Psi=[\Psi_1(r, \varphi),  \Psi_2(r, \varphi)]^{T}$ is the two-component spinor with $\Psi_1(r, \varphi)= e^{im\varphi}\chi_A$ and  $\Psi_2(r, \varphi)=ie^{i(m+1)\varphi}\chi_B$,  where $m=0, \pm 1, \ldots$. This boundary condition requires that $\Psi_{2}/\Psi_{1}=i\tau e^{i\varphi}$ for circular confinement. The two components of the wave function $\chi_A$ and $\chi_B$ correspond to sublattice $\mathcal{A}$ and $\mathcal{B}$. Substituting $\Psi$ into the Dirac equation,  we find the following coupled differential equations:
\begin{equation}\label{1}
v_{F}\displaystyle{\frac{d\chi_B(r)}{dr}}+v_{F}\left(\frac{m+1}{r}+\frac{eBr}{2}\right)\chi_B(r)=E\chi_A, ~~~
-v_{F}\displaystyle\frac{d\chi_A(r)}{dr}+v_{F}\left(\frac{m}{r}+\frac{eBr}{2}\right)\chi_A(r)=E\chi_B, \end{equation}
where we have used the Pauli matrices in cylindrical coordinates
\begin{equation}
\sigma^r=\left(\begin{matrix}0&e^{-i\varphi}\\e^{i\varphi}&0 \end{matrix}\right), ~~~~\\
\sigma^\varphi=i\left(\begin{matrix}0&-e^{-i\varphi}\\e^{i\varphi}&0 \end{matrix}\right), ~~~~\\
\sigma^z=\left(\begin{matrix}1&0\\0&-1 \end{matrix}\right).\end{equation}
From (\ref{1}) we obtain the following differential equations
\begin{equation}\label{wave-r}
\begin{array}{l}
\displaystyle\frac{d^2\chi_A(r)}{dr^2}+\frac{1}{r}\frac{d\chi_A(r)}{dr}-\left(\frac{m^2}{r^2}+\frac{m+1}{l_B^2}-k^2+\frac{r^2}{4l_B^4}\right)\chi_A(r)=0, \\
\displaystyle\frac{d^2\chi_B(r)}{dr^2}+\frac{1}{r}\frac{d\chi_B(r)}{dr}-\left(\frac{(m+1)^2}{r^2}+\frac{m}{l_B^2}-k^2+\frac{r^2}{4l_B^4}\right)\chi_B(r)=0.
\end{array}
\end{equation}
where factor $1/k$ multiplied by equation $\chi_{B}(r)$ is removed and $l_B=(eB)^{-1/2}$ denotes the magnetic length and $k=E/v_{F}$. Equation (\ref{wave-r}) can be further modified as
\begin{equation}
\begin{array}{l}
\displaystyle\frac{d^2\chi_A(s)}{ds^2}+\frac{1}{s}\frac{d\chi_A(s)}{ds}-\left(\frac{m^2}{4s^2}+\frac{m+1}{4s\, l_B^2}-\frac{k^2}{4s}+\frac{1}{16\, l_B^4}\right)\chi_A(s)=0, \\[3mm]
\displaystyle\frac{d^2\chi_B(s)}{ds^2}+\frac{1}{s}\frac{d\chi_B(s)}{ds}-\left(\frac{(m+1)^2}{4s^2}+\frac{m}{4s\, l_B^2}-\frac{k^2}{4s}+\frac{1}{16\, l_B^4}\right)\chi_B(s)=0,
\end{array}
\end{equation}where $s=r^2$.

The authors in [1] found out the solutions with slightly different expressions
\begin{equation}\label{chi-AB}
\begin{array}{c}
\displaystyle\chi_A(s)=\mathcal{C}_ms^{\frac{m}{2}}\exp\left(-\frac{s}{4l_{B}^2}\right)L\left(\frac{k^2l_B^2}{2}-(m+1),  {m}, \frac{s}{2l_{B}^2}\right), \\[4mm]
\displaystyle\chi_B(s)=\frac{1}{kl_B^2}\mathcal{C}_ms^{\frac{m+1}{2}}\exp\left(-\frac{s}{4l_{B}^2}\right)\left[L\left(\frac{k^2l_B^2}{2}-(m+1),  {m}, \frac{s}{2l_{B}^2}\right)+L\left(\frac{k^2l_B^2}{2}-(m+2),  {m+1}, \frac{s}{2l_{B}^2}\right)\right], \end{array}
\end{equation}where $s=r^2$.

Using the boundary condition $\Psi_2(r, \varphi)/\Psi_1(r, \varphi)=i\tau e^{i\varphi}$,  they got the energy expression
\begin{equation}\label{energy}
\left(1-\tau\frac{k l_{B}}{R/l_{B}}\right)L\left(\frac{k^2l_{B}^2}{2}-(m+1), m, \frac{R^2}{2l_{B}^2}\right)+L\left(\frac{k^2l_{B}^2}{2}-(m+2), m+1, \frac{R^2}{2l_{B}^2}\right)=0
\end{equation}

We have to point out that the solutions (\ref{chi-AB}) are incorrect and consequently the energy level equation (\ref{energy}) is flaw. Let us list these crucial mistakes below.

1)Considering the positive and negative quantum number $m$ which determines the behaviors of the wave function near the origin,  the {\it correct} solutions should be written as
 \begin{equation}\label{chi-AB-correct}
\chi_A(s)=\mathcal{C}_ms^{\frac{|m|}{2}}\exp\left(-\frac{s}{4l_{B}^2}\right)\ _1F_1\left(a,  |m|+1,  \frac{s}{2l_{B}^2}\right), ~~~\chi_B(s)=\frac{1}{k\, l_{B}^2}\mathcal{C}_ms^{\frac{|m+1|}{2}}\exp\left(-\frac{s}{4l_{B}^2}\right)\ _1F_1\left(\alpha,  1+|m+1|,  \frac{s}{2l_{B}^2}\right),
\end{equation} where $a=1+(m+|m|)/2-k^2l_B^2/2$ and $\alpha=(1+m+|m+1|)/2-k^2l_B^2/2$. For $m\geq 0$,  we have $a=\alpha=1+m-k^2l_B^2/2$,  but for $m<0$,  we have $a=\alpha=1-k^2l_B^2/2$. This means that $a$ or $\alpha$ is independent of the $m$ for negative $m$.

2)For the {\it unconfined} system,  i.e. in the limits $B\rightarrow 0$ and $R/l_{B}\rightarrow \infty$,  the normalizability requires $a$ or $\alpha$ to be a non-positive integer,  $a=-n=0, -1, -2, \ldots$,  which gives the Landau energy levels
\begin{equation}\label{}
E=\left\{
\begin{array}{ll}
\pm v_{F}\sqrt{[2(n+1)+2m]e\, B}, &~~~m\geq 0, \\[3mm]
\pm v_{F}\sqrt{2(n+1)e\, B}, &~~~m<0,
\end{array}
\right.
\end{equation} which can be unified as
\begin{equation}\label{}
E=\pm v_{F}\sqrt{[2(n+1)+m+|m|]e\, B}, ~~~\hbar=1.\end{equation}
This implies that the Landau levels are independent of the number $m$ for the case $m<0$. Therefore,  the Landau levels given by Eq.(11) of [1] are incorrect.

3) We note that for $a=-n$,  the confluent hypergeometric function $_1F_1(-n, b+1, x)$  reduces to a generalized Laguerre polynomial $L(n, b, x)$. The authors of Ref. [1] made use of this relation to get Eqs. (5) and (6) of Ref. [1]. In fact,  once considering an important identity $L\left(n,  m, x\right)+L\left(n-1,  m+1, x\right)=L\left(n, m+1, x\right)$ we will find that $\psi_{2}(r, \phi)\propto L\left(k^2l_{B}^2/2-(m+1), m+1, r^2/2l_{B}^2\right)$.  Thus the implicit energy equation should have been given by
\begin{equation}\label{}
\tau\frac{k l_{B}}{R/l_{B}}L\left(n, m, \frac{R^2}{2l_{B}^2}\right)=L\left(n, m+1, \frac{R^2}{2l_{B}^2}\right).
\end{equation} Obviously,  it is incorrect for the authors of Ref. [1] to obtain the energy spectrum by using Eq.(6) of Ref.[1]. This is because $n-1$ appeared in $L(n-1, m+1, k^2l_B^2/2)$ if $n=k^2l_B^2/2-(m+1)$ ($m\geq 0$) was taken. This had to make them calculate the energy levels from quantum number $n=1, \ldots, 6$ but not from $n=0, \ldots, 5$ (see subsection C of Sec.II in page 195427-2 of Ref.[1]). Moreover,  only for unconfined systems,  it is possible to substitute $_1F_1(a, b+1, x)$  by $L(n, b, x)$ through taking $a=-n$. However,  for confined system,  i.e. for present graphene quantum dot,  such a substitution is not allowed at all.

\begin{figure}[htb]
\centering
\includegraphics[width=8cm]{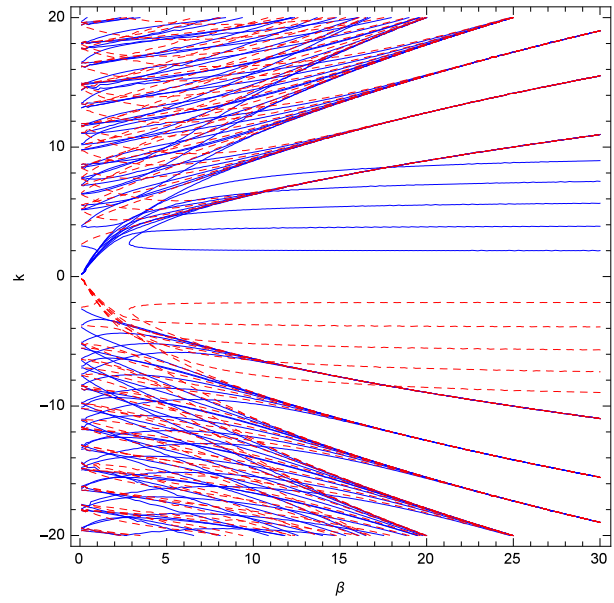}
\caption{(Color online) Reduced energy spectrum of a graphene
quantum dot with reduced size $R =1$ in a magnetic field. The energy levels corresponding to the
$\tau=1$ and  $\tau=-1$ are shown,  respectively,  by the blue solid curves
and the red dashed curves.} \label{F1}
\end{figure}

4)As a result,  the correct energy level equations are given by
\begin{equation}\label{keq}
\begin{array}{l}
\displaystyle\tau\frac{k l_{B}}{R/l_{B}}\, _1F_{1}[a, m+1, R^2/2l_{B}^2]=\, _1F_{1}[a, m+2, R^2/2l_{B}^2], ~~~~m\geq 0.\\[3mm]
\displaystyle\tau\frac{k l_{B}}{R/l_{B}}\, _1F_{1}[\alpha, |m|+1, R^2/2l_{B}^2]=\, _1F_{1}[\alpha, 1+|m+1|, R^2/2l_{B}^2], ~~~~m<0.
\end{array}
\end{equation}
We plot curves of reduced energy spectrum correspond to magnetic field parameter $\beta=\frac{R^2}{2 l_B^2}$  with reduced size $R =1$  for $\tau=\pm 1$ and $m=0, \pm 1, \pm 2, \pm 3, \pm 4$,  respectively,  in Fig. \ref{F1}.

{\Large Acknowledgments}: BJF acknowledges eJDS (ICTP). This work is partially supported by 20150964-SIP-IPN,  Mexico.

\end{document}